\documentstyle[prb,preprint,aps,epsf]{revtex}
\draft

\def\AmS{{\protect\the\textfont2
        A\kern-.1667em\lower.5ex\hbox{M}\kern-.125emS}}

\makeatletter
\def\thepage{1-\@arabic\c@page}
\def\@pnumwidth{2em}
\makeatother 
   \def\<{\langle}
   \def\>{\rangle} 
    \def\one{1\hskip-.37em 1} 
\begin{document}

\title{Electronic conduction in multi-walled carbon nanotubes: Role of
intershell coupling and incommensurability}  \author{Stephan Roche${\
}^{\ddagger}$, Fran\c cois Triozon${\ }^{\dagger}$, Angel Rubio${\ }^{*}$ and Didier Mayou ${\
}^{\dagger}$}

\address{ ${\ }^{\ddagger}$Commissariat \`a l'\'Energie
Atomique, DRFMC/SPSMS, Grenoble, France.\\
  ${\ }^{\dagger}$ LEPES-CNRS, avenue des Martyrs BP166,
38042 Grenoble, France.\\
${\ }^{*}$Departamento de F{\'\i}sica de Materiales, Facultad de
Ciencias Qu{\'\i}micas, Universidad del Pais Vasco/Euskal Herriko
Unibertsitatea, Apdo. $1072, 20018$ San Sebasti\'an/Donostia, Basque
Country, Spain.\\}

\maketitle

\begin{abstract}Geometry incommensurability between weakly
coupled shells in multi-walled carbon nanotubes is shown to be the origin of
unconventional electronic conduction mechanism, power-law scaling of
the conductance, and remarkable magnetotransport and low temperature
dependent conductivity when the dephasing mechanism is dominated by weak
electron-electron coupling. 
\end{abstract}

\vspace{0.2in}

\vfill\eject

Single-walled carbon nanotubes (SWNTs) can be either
metallic of semiconducting depending on their
helicities, i.e. how the graphene
sheet is rolled up\cite{Saito}. 
For weak uniform disorder, the application of the
Fermi golden rule for {\it metallic SWNTs} have
demonstrated $\mu m$ long mean free paths
\cite{Todorov} in agreement with experiments,
clearly pointing towards ballistic
transport\cite{CN-FET,Bachtold2,CN-FET2}.

Multi-walled carbon nanotubes (MWNTs) typically consist of several tenths of
coaxial shells with intershell coupling between neighboring layers. The
understanding of intrinsic properties of MWNTs remain elusive, since different
transport scenario have been reported from ballistic, diffusive and low
resistive, to insulating behaviors\cite{Ebbessen,MWNT-QQ,AB-NT}. An
underestimated feature of MWNT's geometry is that depending on their helical
vector $(n,m)$ (expressed in the basis of the honeycomb lattice), two
consecutive shells are commensurate (resp. incommensurate), if the ratio
between their respective unit cell lengths $T_{\small{(n,m)}}$ along the tube
axis is a rational (resp. irrational) number\cite{Saito}. An illustrative case
is given on Fig.1  where the two double-walled nanotubes $(9,0)@(18,0)$ and
$(9,0)@(10,10)$ are shown. Apparently,  these defect-free structures should
conduct electrons in a similar fashion since they are constituted of metallic
shells with nearly identical diameters. Notwithstanding, by introducing
coupling between shells, the $(9,0)@(18,0)$ remains a periodic system, with a
unit cell of length $T_{\small{(9,0)}}=T_{\small{(18,0)}}\sim 2.15\AA$,
whereas the $(9,0)@(10,10)$ can not be described by a single cell, since 
$T_{\small{(10,10)}}/T_{\small{(9,0)}}=\sqrt{3}$, so that coupling strengths
between nearest-neighbors carbon atoms located at different shells, are not
translationally invariant along the tube axis. The system is therefore
not-periodic and electronic wave-coherence is expected to be qualitatively
different. Incommensurate structures have an increasingly higher probability
to be synthetized than fully commensurate structures, when increasing the
number of inner shells. 

Incommensurability was recently shown to yield anomalous
corrugation properties \cite{Crespi}. Non-ballistic conduction was reported
and implications on magnetotransport discussed\cite{Roche-PRL}.  
Hereafter after defining the hamiltonian together with the
computational method used, several consequences
of the non-ballistic conduction are analyzed for transport coefficients in
MWNTs.

{\it Hamiltonian parameters for the MWNTs}.-Our tight-binding model features one
$p_{\perp}$-orbital per carbon atom, zero
onsite energies, and constant 
nearest-neighbor hopping on each layer n (n.n.), and
hopping between neighboring  layers (n.l.) are defined by~\cite{Saito2}:

{\small $$
{\cal H} = \gamma_{0} \Biggl[\sum_{i,j \hbox{ } n.n.}
|p_{\perp}^{j}\> \<
p_{\perp}^{i}| \Biggr] - \beta \Biggl[\sum_{i,j \in n.l.}
\cos(\theta_{ij})
e^{\frac{d_{ij}-a}{\delta}} |p_{\perp}^{j}\> \<
p_{\perp}^{i}|\Biggr]
$$}

\noindent
where $\theta_{ij}$ is the angle between the $p_{\perp}^{i}$
and $p_{\perp}^{j}$
orbitals, and $d_{ij}$ denotes their relative
distance. The
parameters used here are~: $\gamma_{0}=2.9 eV$, $a=3.34\AA$, 
$\delta= 0.45\AA$. Estimate based on ab-initio calculations for
$\beta$ gives $\beta\simeq\gamma_{0}/8$, but in order to get
insight in  the
effect of $\beta$ on transport properties (since small change in the
interlayer distance yield large variation of the coupling strength), several
values of $\beta$ ($0\leq\beta\leq\gamma_{0}$) have been considered.
Synthetized MWNTs contain typically a few tenth of inner layers, but here the
study is restricted to 2 and 3-walled nanotubes, taking the intershell
distance of $3.4\AA$ as in graphite.

{\it Quantum diffusion in MWNTs}.-We investigate quantum dynamics through the
propagation of wavepackets in long MWNTs with about $1000$
unit cells ($\sim$ half million of carbon atoms). The
time-dependent Schr\"odinger equation is solved numerically by using a
polynomial expansion of the 
evolution operator. It gives us access to the  
diffusion coefficient of a wavepacket $|\psi \>$. If $|\psi \>$ is initially
localized at the center (x=0) of the nanotube, then its diffusivity is simply
$D_{\psi}(t) = L_{\psi}(t)^{2}/t$ with  $L_{\psi}(t) = {\small
\sqrt{\< \psi| (\hat{X}(t)-\hat{X}(0))^{2} |\psi \> }}$ the spatial
spreading of the wavepacket ($\hat{X}$ is the position operator along the tube
axis), and the system size is taken sufficiently large to avoid
boundary effects. For random-phase or energy-filtered
initial states extented to the whole system\cite{TriozonRM}, 
periodic boundary conditions are used for the tube of length
${\cal L}$, and the diffusivity is approximated as $D_{\psi 
}(t) \simeq 4\pi^{2}/{\cal L}^{2}I_{\psi }(t)/t$, where $I_{\psi }(t) = \< 
\psi |\hat{A}^{+}(t)\hat{A}(t)|\psi \>$ and $\hat{A}(t) = \exp(2i\pi 
\hat{X}(t)/{\cal L}) \exp(-2i\pi \hat{X}(0)/{\cal L})- \one $. This
estimation of  $D_{\psi }(t)$ is accurate only for diffusion lengthes
smaller than ${\cal L}$, as checked by finite-size scaling
analysis. The average spreading $L(t)$ and the average diffusion coefficient
$D(t)$ are defined by  $L(t)= \sqrt{\< L_{\psi}^{2}(t)\> } =\sqrt{t D(t)}$,
where $\< \>$ denotes an average over many initial wavepackets.
$D(\tau_{\varphi})$ is the average diffusivity along the
nanotube axis, if at $\tau_{\varphi}$ the electronic wavefunction
looses its phase memory due to some inelastic scattering. The diffusion
coefficient at $\tau_{\varphi}$ are thus connected to the Kubo conductivity
$\sigma=e^{2}\rho D(\tau_{\varphi})$, where $\rho$ is the density
of states. This approach provides a good qualitative picture of wavepacket
propagation in MWNTs constituted of conducting shells, with energies away from
the charge neutrality point. Experimental studies demonstrate that chemical
potential away from the charge neutrality point can be obtained upon
doping\cite{collins,dop2,kruger}. Kr\"uger et al.\cite{kruger} obtained
variations of the Fermi energy of the order of $\sim 0.3eV-0.5eV$,
corresponding to $10-15$ conducting channels instead of $2$ per metallic layer.

{\it Anomalous conduction in incommensurate
MWNTs}.-For disorder-free commensurate systems, a wavepacket initially
localized in the outermost shell of the MWNTs, quickly transfer its weight
 to inner shells, due to intershell coupling. By
following its spreading properties, we found that the wavepacket propagation
is ballistic along the tube-axis over large distances ($\mu$m), i.e. $L(t)\sim
vt$ similarly to the case of perfect metallic single-walled nanotube.

However, for the incommensurate case, the transfer of the electronic
wavepacket is followed by a non ballistic propagation given
by $L(t) \sim {\cal A}t^{\eta}$. The coefficient
$\eta$ is found to decrease from $\sim 1$ to $\sim 1/2$ by increasing the number of coupled
incommensurate shells\cite{Roche-PRL}. On Fig.2, ballistic versus non ballistic conduction
mechanisms are illustrated in the defect-free incommensurate
double-walled nanotubes $(9,0)@(18,0)$ and
$(9,0)@(10,10)$. The case $(6,4)@(10,10)@(17,13)$ is also shown to demonstrate
the effect of increasing the "strength" of incommensurability on the
wavepacket propagation. The interlayer coupling parameter taken here is given
by $\beta=\gamma_{0}/3$ which is slightly larger than the expected ab-initio
value. By increasing $\beta$, a reduction of 
power-law exponent $\eta$ is obtained, similarly to what is found by increasing
the number of coupled incommensurate shells, for a fixed value of $\beta$.

{\it Magnetotransport properties}.-The effect of magnetic field and disorder
when charges propagate predominantly on the outermost shell has been analyzed in 
\cite{Roche-PRL}. Here, we focus on the magnetotransport in incommensurate MWNTs if the magnetic
field is applied parallel to the tube axis. Magnetic field dependent diffusion
coefficients of electronic wavepackets are given on Fig.3., for
incommensurate 2-wall $(9,0)@(10,10)$ (main frame) in the power-law diffusion
regime (with two values of $\beta=\gamma_{0}/3,\gamma_{0}$).
$\Phi_{0}$-periodic oscillations and positive magnetoresistance at low field
are observed, similar to what is predicted for the ballistic regime in the
SWNT case\cite{Ando}. Note that the Aharonov-Bohm oscillations for the
commensurate $(9,0)@(18,0)$ are still $\Phi_{0}$-periodic, but with much
smaller amplitude (not shown here). By increasing $\beta$, the coupling
strength between walls, the magnetoconductance is reduced without modification
of the oscillatory behavior.

Differently, for the incommensurate nanotube $(6,4)@(10,10)@(17,13)$ (inset of
Fig.3), in which diffusive-like propagation takes place, there is evidence for
negative magnetoresistance at low field and $\Phi_{0}/2$-periodicity of
$D(\tau_{\phi},\Phi)$. From the saturation of diffusivity, we deduce some
effective elastic mean free path $\tilde{l}_{e}$, that turns out to be
slightly smaller that the outer shell circumference (for
$\beta=\gamma_{0}$). By considering a reduced value of the
coupling parameter ($\beta=\gamma_{0}/3$-not shown here), despite the fact
that the effective mean free path becomes larger than the outershell
circumference, still negative magnetoresistance at low field and
$\Phi_{0}/2$-periodic oscillation persist. These results confirm that
magnetotransport properties of MWNTs are very sensitive to the geometry, the
number of shells carrying current, and the hamiltonian parameters. In
disordered systems, the basic scheme is that of ballistic electrons scattering
on random impurities. The $\Phi_{0}/2$ periodicity showed here, also results
from quantum interferences of the electronic pathways around the cylinder
wrist. However the usual scheme of interfering backscattered electronic
trajectories is not strictly applicable for incommensurate disorder-free
MWNTs, as confirmed by a study of the probability of return to the origin of
wavepackets\cite{TR}. Fujiwara et al.\cite{AB-jap} observed a surprising
$\Phi_{0}/3$-oscillation of magnetoconductance and gave an interpretation in
terms of superposition of phase shift due to spectral effects, assuming three
internal tubes with different helicity (same situation as in 
$(6,4)@(10,10)@(17,13)$). The effect of magnetic field on density of states
(DoS) has been discussed theoretically\cite{Ando2} and should also be
considered to fully understand the magnetotransport in SWNTs and moderately
large diameter MWNTs.

{\it Electron-electron interactions and temperature dependence of inelastic
scattering times and conductivity}-.We focus now on the combined effect of
 anomalous propagation of wavepackets
(ASWP) together with a dephasing mechanism dominated
by electron-electron interactions. Recently, the influence of ASWP on
frequency and temperature dependences of inelastic
scattering times and conductivities, in systems close
to a metal-insulator
transition\cite{ADTS1}, in aperiodic incommensurate
and quasiperiodic
structures\cite{AD2TS,AD3TS,Mayou} has been
demonstrated. 

The effect of electron-electron
interaction in transport
properties of MWNTs can be addressed within two limits.
The strong coupling case is
related to the existence of a Luttinger liquid phase,
which has been
discussed theoretically recently for metallic
commensurate MWNTs \cite{13}. The
effect of incommensurability on Luttinger liquid was
shown in other
context to develop new phase diagram\cite{14}, result
that could be valuable
in the context of incommensurate carbon nanotubes. In the
weak coupling limit, the electron-electron interaction acts as the main
dephasing mechanism and monitor the temperature dependences
of inelastic scattering times and conductivity. 

The interference pattern between two waves   
$\Psi_{1}(t)$ and $\Psi_{2}(t)$ which propagate coherently within the system
is obtained from the evaluation of the relative phase difference accumulated
in time. In the limit of low
temperature, the inelastic coupling with other electrons, which produces a loss
of phase coherence, can be expressed by an external time-dependent
potential $V(L_{\Psi}(t),t)$, whose effect is to superimpose new phase
factors 

$$\langle e^{i\phi}\rangle = \langle e^{i(\int
V(L_{\Psi}(t),t)dt/\hbar)}\rangle= \int
{\cal P}(\phi) e^{i\phi}d\phi$$

\noindent
with $\phi$ defined by a probability distribution ${\cal P}(\phi)$. As the
potential is time-dependent, the initial phase interference pattern is reduced
in amplitude since the phase factors are modified by uncorrelated phase shifts. 
The loss of phase coherence has been shown to be driven by
the increase of phase uncertainty between the superimposed phases accumulated
by the two non-interacting waves $\Psi_{1}(t)$ and $\Psi_{2}(t)$ and related
(under certain approximation, see \cite{TSEE2}) to
$\langle\delta\phi_{1-2}^{2}\rangle=\langle \phi_{1-2}^{2}\rangle- \langle
\phi_{1-2}\rangle^{2}$ ($\phi_{1-2}=\int dt 
V(L_{\Psi_{1}}(t),t)-V(L_{\Psi_{2}}(t),t)$) with {\small
$V(L_{\Psi}(t),t)=-\frac{e}{c} \frac{dL_{\Psi}(t)}{dt}\cdot {\cal 
A}(L_{\Psi}(t),t)$}, where ${\cal A}(L_{\Psi}(t),t)$ is the potential vector.
The inelastic phase coherence time $\tau_{\varphi}$ defines the limit of
coherent phase interference and thus comes out from\cite{TSEE2}

$$\langle \delta\phi_{1-2}^{2}\rangle\simeq e^{2}k_{B}T\int_{0}^{\tau_{\varphi}}dt\mid   
L_{\Psi_{1}}(t)-L_{\Psi_{2}}(t)\mid^{2-d}\simeq
1$$ 

\noindent
In a diffusive medium, $\mid
L_{\Psi_{1}}(t)-L_{\Psi_{2}}(t)\mid\sim
t^{1/2}$, which triggers $\tau_{\varphi}^{-1}\sim T^{2/3}$,
behavior clearly
identified experimentally at low temperature ($T<4K$)
in metallic wires\cite{TSEE3}.

In defect-free incommensurate MWNTs, ASWP implies a power-law dependent
diffusivity $D(\tau_{\varphi})\sim \tau_{\varphi}^{2\eta-1}$, which thus
results in a $\tau_{\varphi}^{-1}\sim T^{1/(1+\eta)}$ power-law for
the inelastic scattering times. Subsequently, an anomalous
temperature dependence of the conductivity follows 

 $${\displaystyle \sigma(T)\sim T^{\displaystyle
\frac{(1-2\eta)}{(1+\eta)}}}$$

In the main frame of Fig.4 such behavior is exemplified for different
exponents $\eta$. One notes that a temperature
independent conductivity is found for the limit
$\eta\to 1/2$ (diffusive-like motion), and the decrease of
conductivity with temperature may be completely opposite for subdiffusive
regime ($\eta<1/2$). Although in our calculations $\eta$ remains above
$1/2$, such possibility, often found in
quasiperiodic systems, can not be fully discarded for more complex situations.
Recently Natelson et al. \cite{Natelson} have reported a continuous change of
the power law exponent of the inelastic scattering times $\tau_{\varphi}$ by
varying the geometry of a quantum wire, i.e. by tuning quantum interferences.
Here, we have ascribed similar behavior to the anomalous propagation steming
from incommensurability in MWNTs. Along the same lines, one also notes that a
study of the frequency dependent conductivity could be an
interesting way to unveil anomalous quantum diffusion\cite{Mayou}. Frequency dependent study of
electronic impedance has been recently demonstrated to be relevant for carbon
nanotubes\cite{Ajayan}.

{\it Anomalous scaling of the conductance}-.The relation between the electronic conductivity
and the conductance allows to anticipate an anomalous length dependence
of the Landauer conductance. Indeed, in a non-ballistic
conduction, electrons propagate according $L(t)\sim {\cal A}t^{\eta}$, with
${\cal A}$ and $\eta$ depending on the MWNT geometry but also the number of
shells participating in transport ($N_{c}$). The conductance is related to the
conductivity through $G(L)\sim\sigma(L)/L$. By defining $D_{nb}(L)$ (resp.
$D_{b}(L)$) the value of the non-ballistic diffusion coefficient (resp.
ballistic diffusion coefficient for a fully commensurate metallic
MWNT with $N_{c}$ shells), then the conductance at scale $L$ writes   

  \begin{eqnarray}
  \frac{G_{nb}(L)}{2N_{c}e^{2}/h} &\sim &
\frac{D_{nb}(L)}{D_{b}(L)}\nonumber\\
G_{nb}(L)&\sim &\frac{2N_{c}e^{2}}{h}\times\alpha\times
L^{(\eta-1)/\eta}  \nonumber     
\end{eqnarray}

\noindent
with a factor $\alpha\sim ({\cal A}/v_{F})^{1/\eta}$ that will depend on the
particular system and chemical potential considered.  Indeed, if the chemical potential is close
to the charge neutrality point, fewer conducting channels will be available
because of the presence of semiconducting shells weakly conducting at such
energy. The ballistic case $\eta=1$ corresponds to $G\sim
\frac{N_{c}e^{2}}{h}$, that is a length-independent conductance, whereas in the
diffusive limit, the conductance scales as $\sim 1/L$ (physically related with
the distance between electrodes). To test such physical effect, experiments
should be carried out on highly purified MWNTs with typical diameter of $\sim
10-20nm$, since it will allow to tune the participation in transport of more
than 10 shells, under reasonably weak chemical doping ($\leq 1\%$), or
directly by varying the applied voltage $V_{bias}$ (the number of shells
$N_{C}$, participating in conduction being a function of $V_{bias}$).
Furthermore, the study of the conductance with the distance between leads
probing the sample requires {\it perfect ohmic contacts} between electrodes and
the nanotube.

{\it Conclusions}-.Incommensurability in MWNTs is a geometrical degree of
freedom at the origin of {\it non ballistic conduction}. The
redistribution of the wavepacket in the whole object is followed by a specific
 multiple wave scattering that jeopardize ballistic conduction. All transport
coefficients should be sensitive to such effects, especially the temperature
dependence of inelastic scattering times and conductivity driven by
electron-electron interactions at low temperature, as well as the
length-scaling of the Landauer conductance. Experiments should be performed in
conjunction with theoretical calculations of the relevant microscopic
parameters (${\cal A},\eta$) that depend on the geometry of the MWNT as well
as the number of shells participating in conduction.

\noindent 

Acknowledgments: Financial support from {\small NAMITECH
[ERBFMRX-CT96-0067(DG12-MITH)], DGES (PB98-0345), COMELCAN(
HPRN-CT-2000-00128), JCyL (VA28/99)} and {\small C$^4$} are acknowledged. S.R. is indebted
to Riichiro Saito and H\'el\`ene Bouchiat for stimulating discussions.

\vfill\eject

\vfill\eject
\noindent 
{\bf Figure captions}:

\vspace{20pt}
\noindent
{\bf Figure 1}: Geometrical representation of $(9,0)@(18,0)$(right) and
$(9,0)@(10,10)$(left) double-walled carbon nanotubes {\bf ASK me by e-mail !}
at {\it sroche@cea.fr}.

\vspace{20pt}
\noindent
{\bf Figure 2}: Main Frame: Averaged diffusion coefficient (arb. unit) for
$(9,0)@(18,0)$ and $(9,0)@(10,10)$ with $\beta=\gamma_{0}/3$. Inset:
Avergared diffusion coefficient for the incommensurate MWNT
$(6,4)@(10,10)@(17,13)$. One notes that $T_{\small{(6,4)}} =
3\sqrt{19}a_{\small cc}$, $T_{\small{(10,10)}} = \sqrt{3}a_{\small cc}$, and
$T_{\small{(17,13)}}=3\sqrt{679}a_{\small cc}$ where $a_{\small cc}=1.42\AA$
is the interatomic distance between carbon atoms.

\vspace{20pt}
\noindent
{\bf Figure 3}: Main frame : Magnetic field dependence of the diffusion
coefficient for the incommensurate double-walled $(9,0)@(10,10)$ for two values
of $\beta=\gamma_{0}/3$ (solid line, right y-axis) and $\beta=\gamma_{0}$
(dashed line, left y-axis) evaluated at the same time $\tau_{\phi}$. Inset :
Case for $(6,4)@(10,10)@(17,13)$ at two different times
$\tau_{\phi}$($\hbar/\gamma_{0}$ unit) for $\beta=\gamma_{0}$.

\vspace{20pt}
\noindent
{\bf Figure 4}:Main Frame: Scaling of the temperature dependence of the
conductivity with decreasing exponent $\eta$. Inset: Length dependence of the
conductance for the same parameters. Bold solid lines are for $\eta\sim 1$, and
bold dashed lines for $\eta\sim 1/2$, whereas other solid lines interpolate
between 1 and 1/2.

\vfill\eject

   \begin{figure}[htbp]
   \epsfxsize=15cm
   \centerline{\epsffile{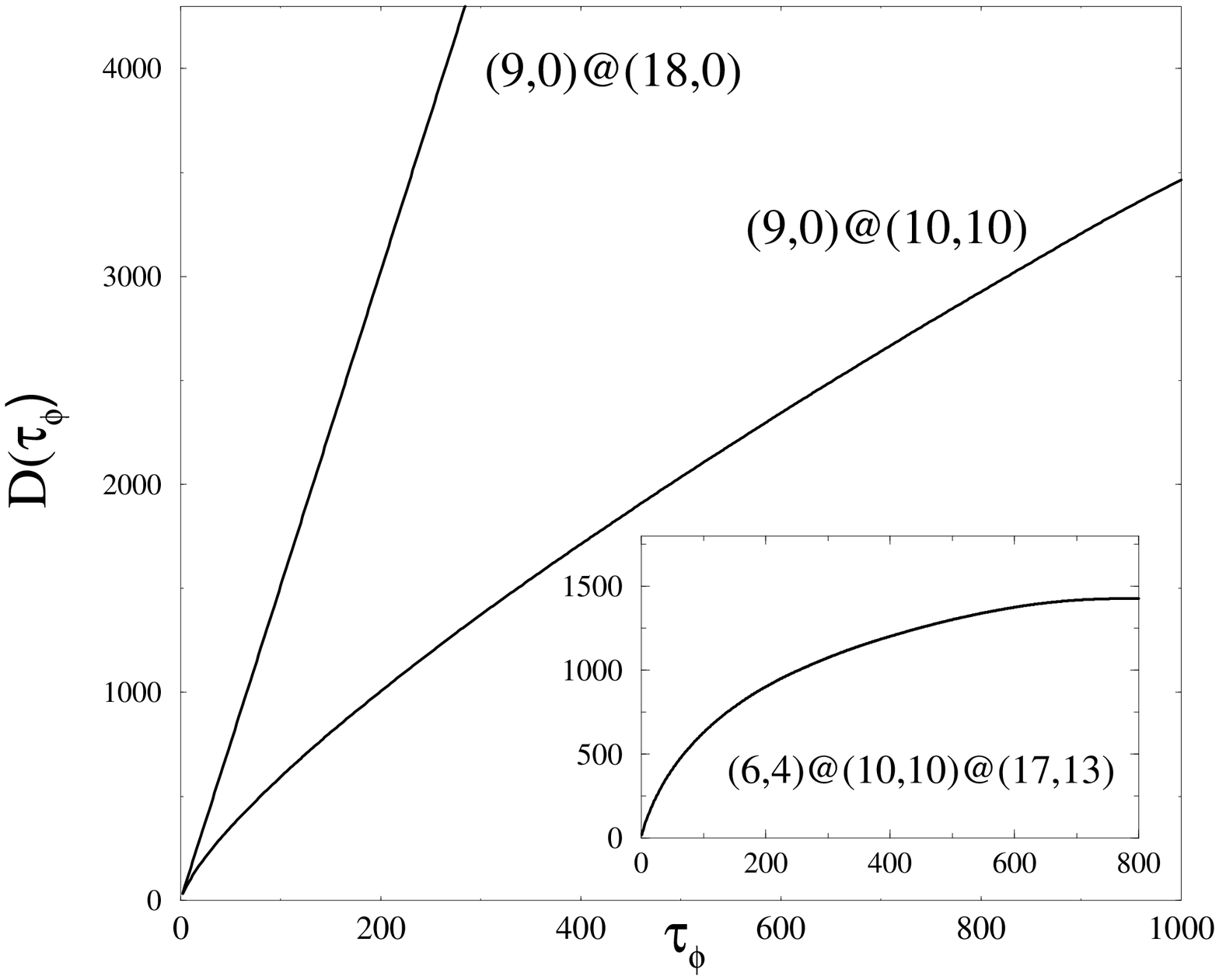}}
    \end{figure}

\begin{figure}[htbp]
\epsfxsize=15cm
\centerline{\epsffile{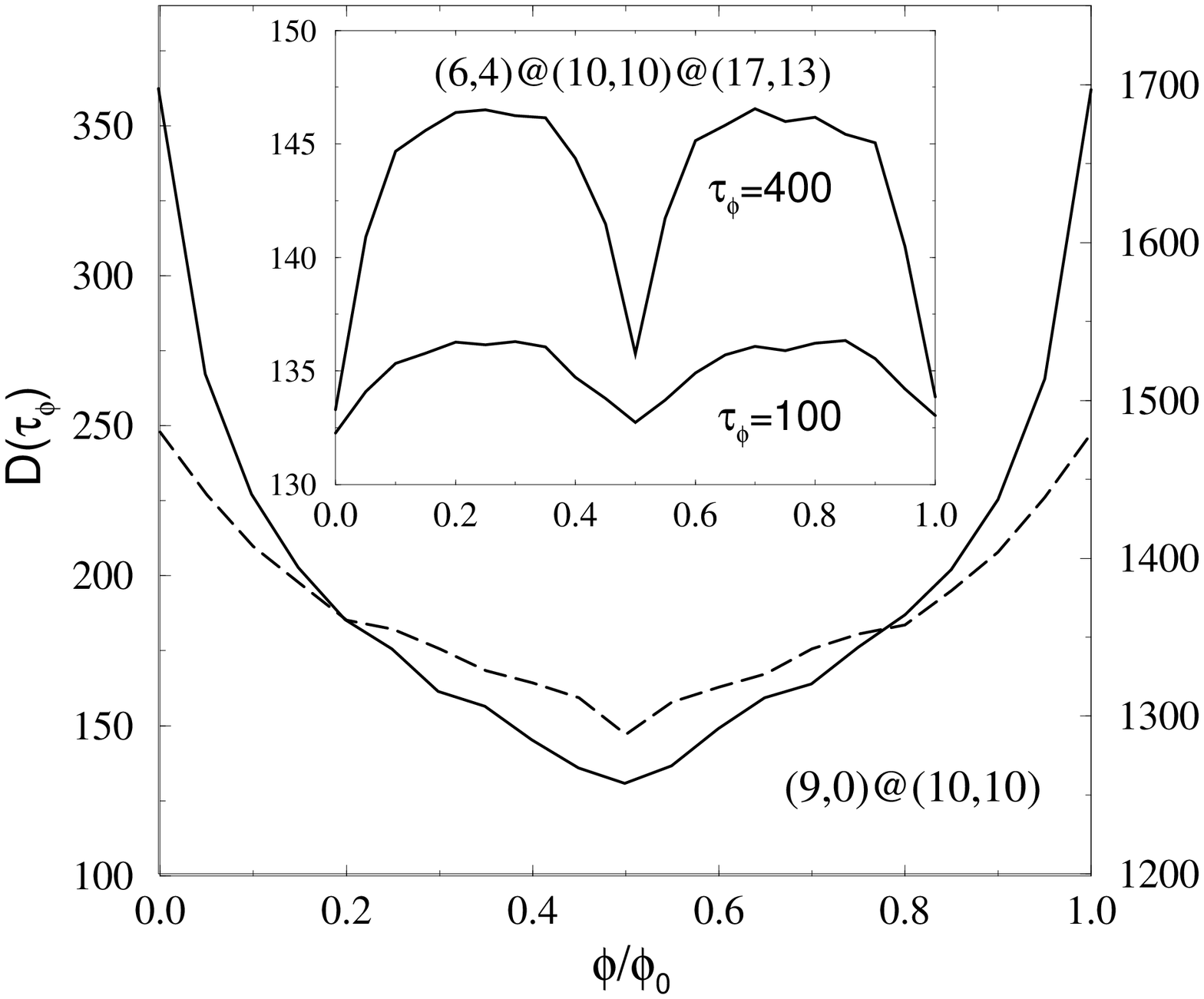}}
\end{figure}

   \begin{figure}[htbp]
   \epsfxsize=15cm
   \centerline{\epsffile{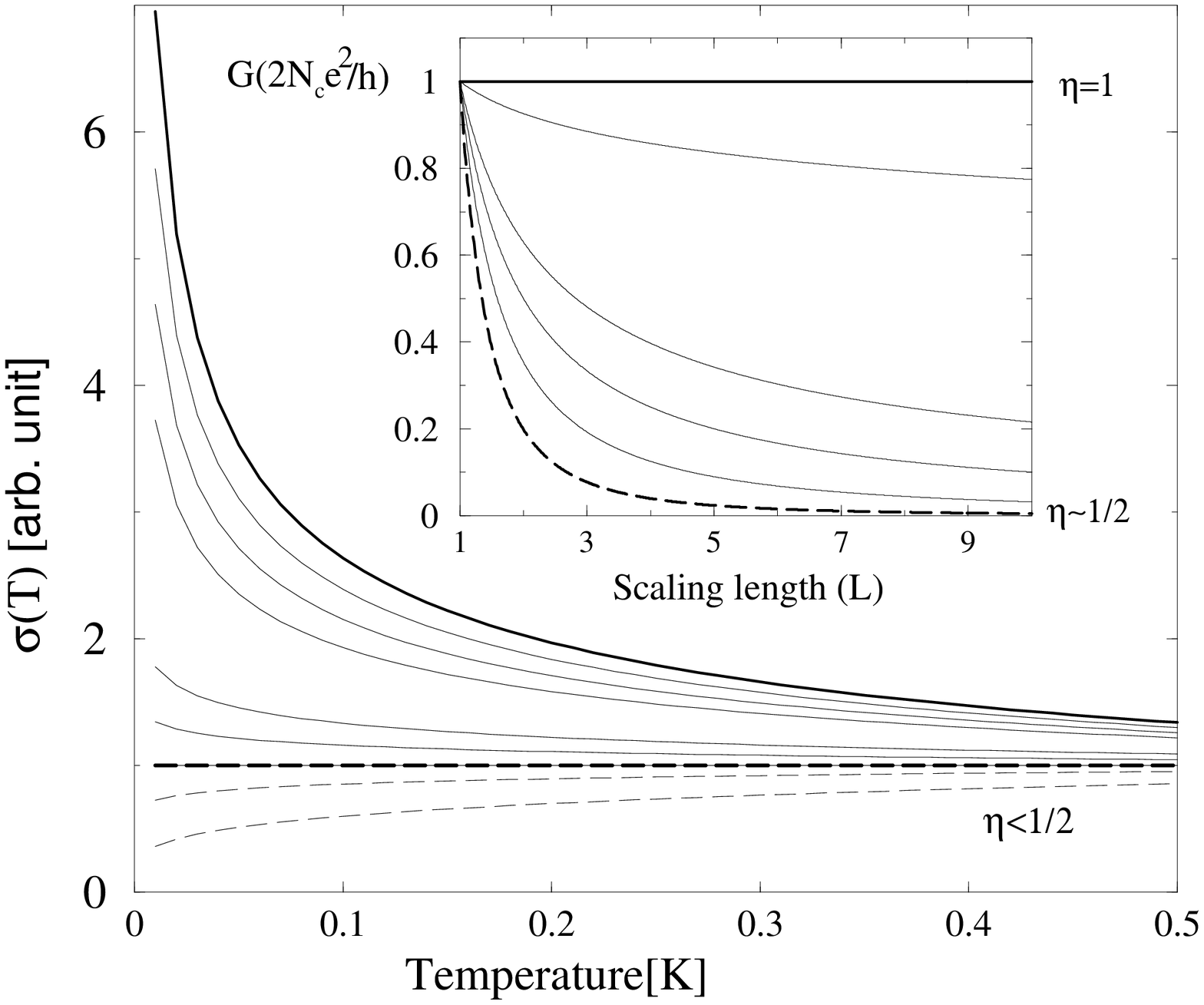}}
    \end{figure}

   \end{document}